\documentclass{astlb}

\usepackage[utf8]{inputenc}
\usepackage[english]{babel}

\usepackage{mathptmx}
\usepackage{txfonts}
\usepackage{pdflscape}
\usepackage[T1]{fontenc}
\usepackage{ae,aecompl}
\usepackage{changes}

\usepackage{graphicx}
\usepackage[hyperfootnotes=false]{hyperref}

\newcommand{\igr}{IGR J21343+4738}
\newcommand{\nus}{NuSTAR}

\DeclareUnicodeCharacter{2212}{-}

\sloppypar
\begin{document}
\journalinfo{2022}{48}{12}{798}[805]
\title{Study of the X-ray Pulsar IGR J21343+4738 based on NuSTAR, Swift, and SRG data}

\author{A.S. Gorban\email{gorban@cosmos.ru}$,^{1, 2}$
S.V. Molkov$,^{1}$ A.A. Lutovinov$,^{1}$ A.N. Semena$,^{1}$ 
\\
$^{1}$Space Research Institute of the Russian Academy of Sciences, Profsoyuznaya 84/32, 117997 Moscow, Russia\\
$^{2}$Higher School of Economics, Myasnitskaya 20, 101000 Moscow, Russia\\
}

\shortauthor{Gorban \etal}

\shorttitle{Study of the X-ray Pulsar IGR J21343+4738} 
\submitted{15.11.2022}

\begin{abstract}
We present the results of our study of the X-ray pulsar IGR~J21343+4738 based on NuSTAR, Swift, and SRG observations in the wide energy range 0.3---79 keV. The absence of absorption features in the energy spectra of the source, both averaged and phase-resolved ones, has allowed us to estimate the upper and lower limits on the magnetic field of the neutron star in the binary system,  $B<2.5\times10^{11}$~G and $B>3.4 \times 10^{12}$~G, respectively. The spectral and timing analyses have shown that IGR~J21343+4738 has all properties of a quasi-persistent X-ray pulsar with a pulsation period of $322.71\pm{0.04}$~s and a luminosity $L_{x} \simeq3.3$ $\times10^{35}$~erg~s$^{-1}$. The analysis of the long-term variability of the object in X-rays has confirmed the possible orbital period of the binary system $\sim 34.3$~days previously detected in the optical range.


\rule{8cm}{1pt}\\
\end{abstract}

\section*{Introduction}

Studying Be/X-ray binaries (BeXRBs) during both outbursts and low states is of great interest, allowing to estimate the physical parameters of the binary systems \citep[see, e.g.,][]{2011Ap&SS.332....1R, 2019NewAR..8601546K, 2022arXiv220414185M}. This type of binaries is a system that consists of a massive, rapidly rotating Be star exhibiting (Balmer and Paschen) hydrogen emission lines in the spectrum and a neutron star, the accretion onto which is the main source of X-ray emission in the system. 
Some of the BeXRBs exhibit outbursts during which the luminosity of the sources can reach the Eddington limit for a neutron star  $L\simeq10^{38}$ erg s$^{−1}$ or more. 
There is also a subclass of BeXRBs that are characterized by a persistent weak X-ray emission and that do not show a pronounced activity. 
The luminosity of such systems does not exceed $\sim 10^{35}$ erg s$^{-1}$. Persistent X-ray binaries have long pulsation periods \citep[for example,][]{2012ApJ...761...49L, 2012MNRAS.421.2407T, 2017ApJ...847...44Q}.

The X-ray pulsar IGR J21343+4738 was discovered by the INTEGRAL observatory during its observations of the Galactic plane in Cygnus \citep{2007A&A...475..775K, 2007yCat..21700175B}. In December 2006 the source IGR J21343+4738 was observed by the Chandra observatory \citep{2008A&A...487..509S}, which allowed it to be accurately localized and its optical counterpart to be determined. The optical spectrum of this source exhibited intense H~I and He~I absorption lines, suggesting a B-type star \citep{2008A&A...487..509S, 2008AstL...34..653B}. Later, \cite{2014A&A...561A.137R} established that the optical counterpart of IGR~J21343+4738 is a B1IVe Be star that has a magnitude V = 14.1 and is at a distance  $\sim\ 8.5$~kpc. The same authors noted that a deep central absorption between the two H$\alpha$ peaks is also present in the optical spectrum of the Be star.

In November 2013 the source was observed by the XMM-Newton observatory. The source was in quiescence with a flux $\simeq1.4 \times 10^{−11}$ erg cm$^{−2}$ s$^{−1}$ in the 0.2–12 keV energy band. A timing analysis of the XMM-Newton data revealed pulsations with a period $\simeq$320~s \citep{2014A&A...561A.137R}.

The orbital parameters of the binary system were estimated by \cite{2021mobs.confE..25N} based on long-term photometric and spectroscopic observations with the Russian–Turkish RTT-150 telescope. These authors found two possible orbital periods of the neutron star around its companion, approximately 34.26 and 160.8 days, and the corresponding eccentricities, 0.36 and 0.38.

In December 2020, during the second sky survey, the Mikhail Pavlinsky ART-XC telescope onboard the SRG observatory recorded a significant increase in the flux from IGR~J21343+4738 \citep{2020ATel14247....1S}, $\sim 5 \times 10^{-11}$ erg/cm$^2$/s in the 4–12 keV energy band, compared to the flux detected in the previous sky survey $\sim 1.3 \times 10^{-11}$ erg/cm$^2$/s. The observations of the source in a wide energy range by the NuSTAR and Swift observatories followed next.

In this paper we perform a detailed analysis of the data from all three observatories to measure the timing and spectral characteristics of the source IGR~J21343+4738 and to estimate its physical parameters.


\section*{OBSERVATIONS AND DATA ANALYSIS}
\label{sec:obs}

The source IGR~J21343+4738 was observed by the NuSTAR observatory on December 17, 2020, (MJD 59200) with an exposure time of about 27 ks (ObsID 90601339002). To investigate the pulsar in a wide energy range, it was observed simultaneously by the NuSTAR \citep{2013ApJ...770..103H} and Neil Gehrels Swift \citep{2004ApJ...611.1005G, 2005SSRv..120..165B} observatories. The data of these observatories were taken from the HEASARC archive (http://heasarc.gsfc.nasa.gov/).

\nus\ (Nuclear Spectroscopic Telescope ARray) -- is an observatory sensitive to hard X-ray emission in the energy range from 3 to 79 keV. The observatory consists of two identical co-aligned X-ray telescopes (FPMA and FPMB) with a spectral resolution of 0.4 keV at 10 keV and about 0.9 keV at 60 keV \citep{2013ApJ...770..103H}. The data were processed with the standard NuSTARDAS v1.9.7 software provided as part of the HEASOFT v6.30.1 package with CALDB v20220426 calibrations. To obtain the energy spectra of the source and its light curves, the data were processed with the NUPIPELINE and NUPRODUCTS procedures. The data from each of the two modules for the source were chosen from circular regions of radius  50\arcsec, centered on the source. The background spectra and light curves were extracted from a region of radius 70\arcsec~located on the same chip away from the source. To expand the investigated energy range, we also used the spectral observations of IGR~J21343+4738 with Swift/XRT in the 0.3–10 keV energy band. The observations were carried out simultaneously with the NuSTAR observatory with an exposure time of 1.6 ks. The energy spectra were obtained by the online services \citep[see][]{2009MNRAS.397.1177E}  provided by the UK Swift Science Data Centre at the University of Leicester\footnote{\url{http://www.swift.ac.uk/user\_objects/}}. The observations were carried out in the photon counting (PC) mode. The data for the source were extracted from a circular region of radius 47\arcsec, centered on the source. All energy spectra of the source were binned in energy by 25 counts per bin to use the $\chi^2$ statistic for our spectral analysis. The spectra were fitted with the {\sc XSPEC} v12.12.1 package \citep{1999ascl.soft10005A}.

The source IGR~J21343+4738 was observed several times during the Spectrum-Roentgen-Gamma \citep[SRG, see][]{2021A&A...656A.132S} all-sky survey. In this paper we used data from the Mikhail Pavlinsky ART-XC telescope \citep{2021A&A...650A..42P} — an X-ray telescope with grazing incidence optics that operates in the hard X-ray energy range 4–30 keV and allows the positions of sources in the sky to be determined with an accuracy of $\approx15"$, and their timing and spectral analyses to be performed. From December 2019 to March 2022 the telescope executed an all-sky survey program during which each position in the sky was observed every six months. For four observations of IGR~J21343+4738 performed during the all-sky survey the flux in the 4–12 keV energy band was measured using standard data processing procedures from the ARTPRODUCTS V0.9 package with CALDB 20200401 calibration data. The fluxes were extracted from a 2\arcmin~aperture. Since the effective exposure time is short and the number of recorded photons is limited, it is difficult to perform detailed timing and spectral analyses.

\section*{RESULTS}
\label{sec:time}

\subsection{Timing Analysis of the Emission from IGR~J21343+4738}

To properly determine the timing characteristics of the source, the photon arrival times were first corrected for the Solar System barycenter using the standard barycorr tool from the HEASOFT package. Since there are no known orbital parameters, no correction for the orbital motion was made. The background was subtracted from the NuSTAR light curves with a resolution of 0.1 s and then the light curves of the two NuSTAR modules were combined using the lcmath tool (XRONOS v6.0). The pulsation period was determined using the epoch-folding technique implemented in the efsearch tool from the HEASOFT package. The pulsation period determined from the NuSTAR data was 322.71 $\pm$ 0.04 s. The error in the period was estimated using Monte Carlo simulations of the light curve \citep[for more details, see][]{2013AstL...39..375B}.

\begin{figure*}[!t]
	\center{\includegraphics[scale=0.7]{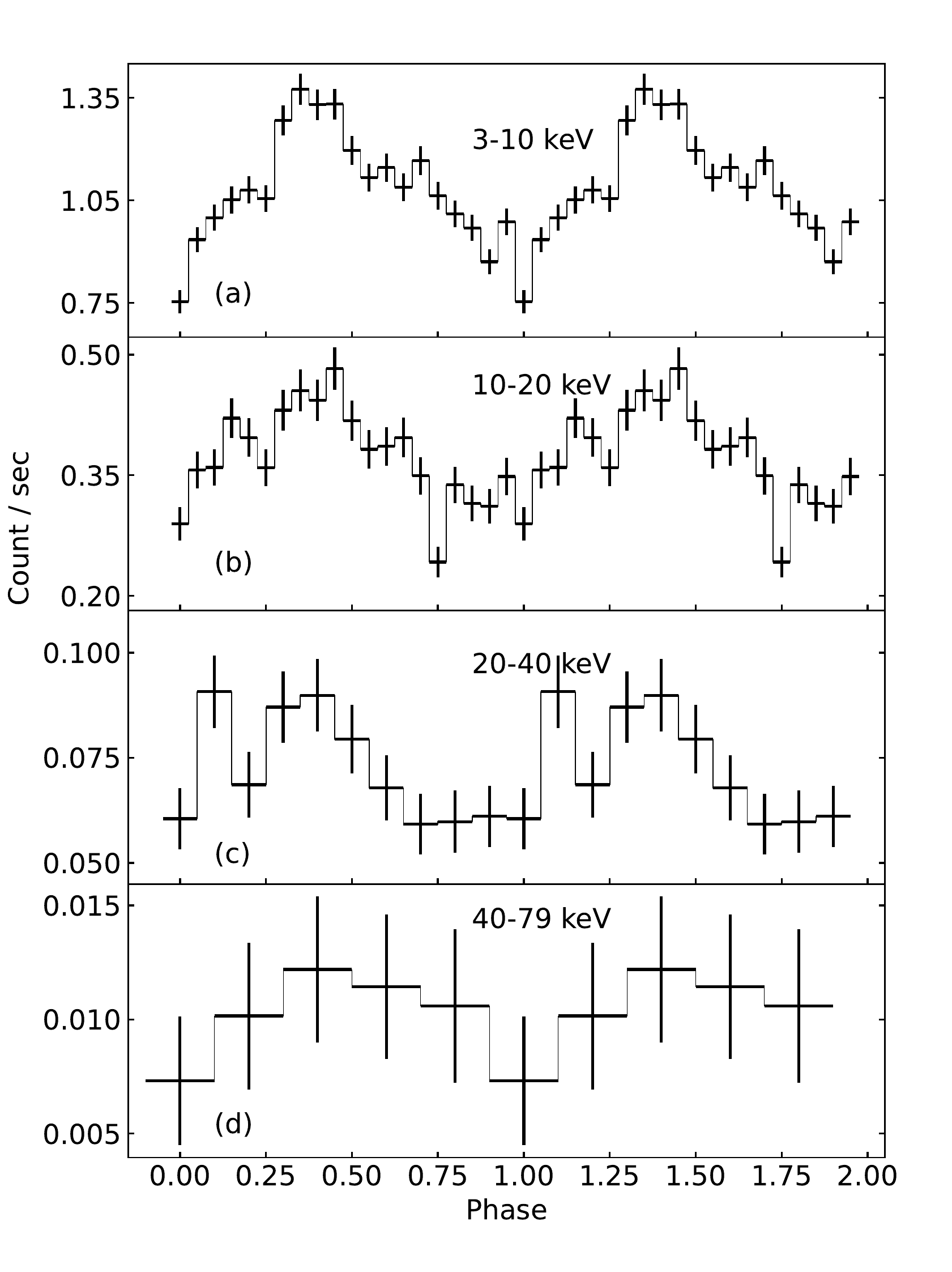}}
	\caption{The pulse profiles of IGR~J21343+4738 in four energy bands from the NuSTAR data in December 2020 (ObsID 90601339002).} 
	\label{fig:pimp}
\end{figure*}

We have researched the pulse profile in various energy bands: 3–10, 10–20, 20–40, and 40–79 keV. The light curves were constructed from the NuSTAR data and were folded with the previously determined period of 322.71 s. Figure 1 presents the pulse profiles for the source. It is clearly seen that the profile hardly changes with energy. A primary peak is recorded at phases 0.3–0.5, which is preserved up to 40 keV. A second small peak is also observed up to 40 keV near phase 0.15–0.2. It can be seen that above 40 keV there is no significant pulsed signal, which is most likely due to the low flux from the source and, as a consequence, the shortage of statistics in the hard energy channels. A similar pulse shape was also observed at softer energies based on the XMM-Newton observations \citep{2014A&A...561A.137R}.

\begin{figure*}[!t]
\centering
\includegraphics[scale=0.8]{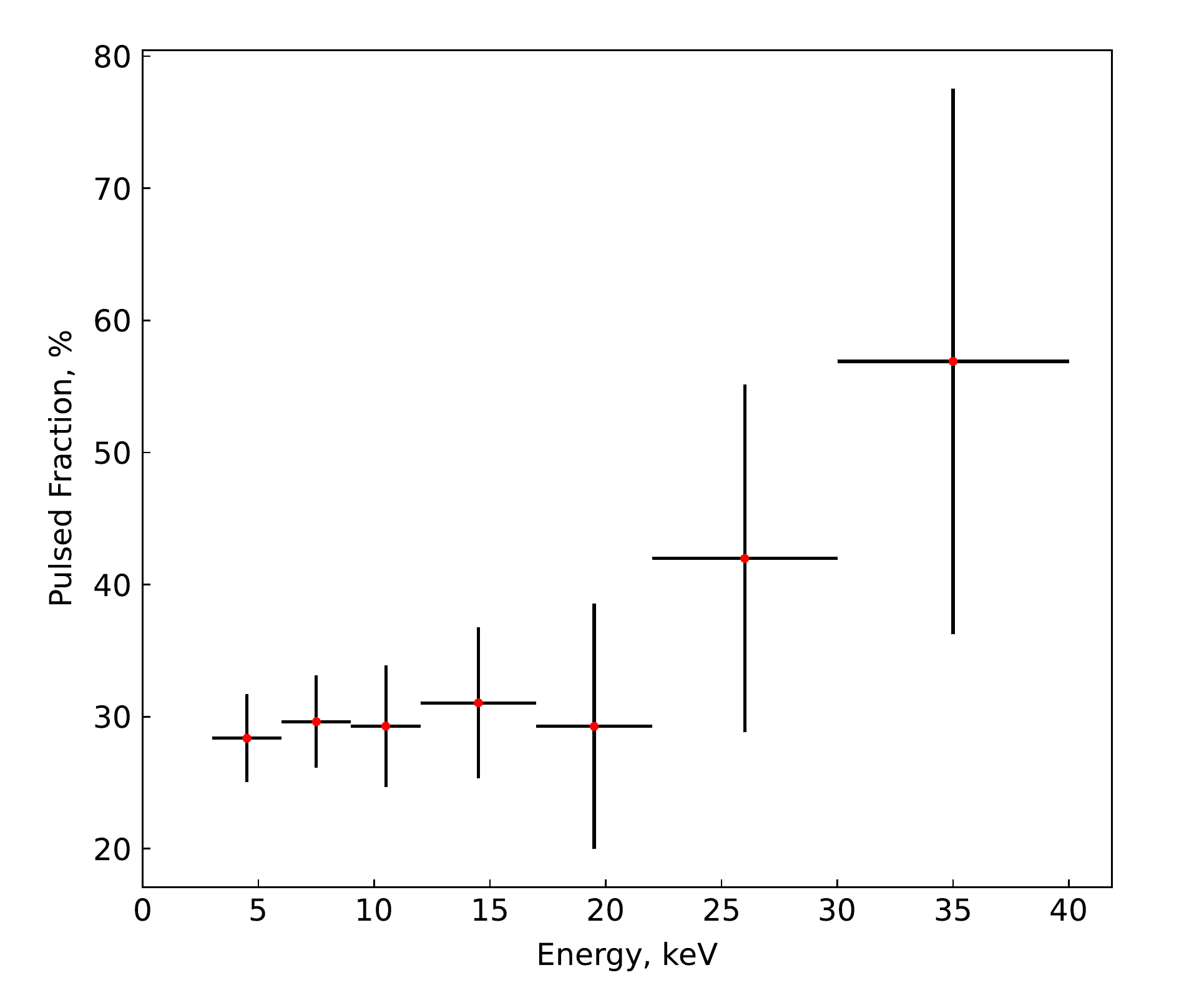}\caption{Pulsed fraction for IGR~J21343+4738 versus energy based on the NuSTAR data (ObsID 90601339002).} 
	\label{fig:pf}
\end{figure*}

The pulsed fraction was calculated as the ratio 
($F_{max}-F_{min}$)/($F_{max}+F_{min}$),
where $F_{max}$ and $F_{min}$ -- are the maximum and minimum fluxes in the background-corrected pulse profile in the chosen energy band consisting of 16 bins. The derived dependence  (Fig. \ref{fig:pf}) demonstrates an increase in the pulsed fraction with energy, which is typical for X-ray pulsars \citep{2009AstL...35..433L}.

\subsection{Spectral Analysis of the Source \igr\ }

\begin{figure*}[!t]
	\centering{\includegraphics[scale=0.82]{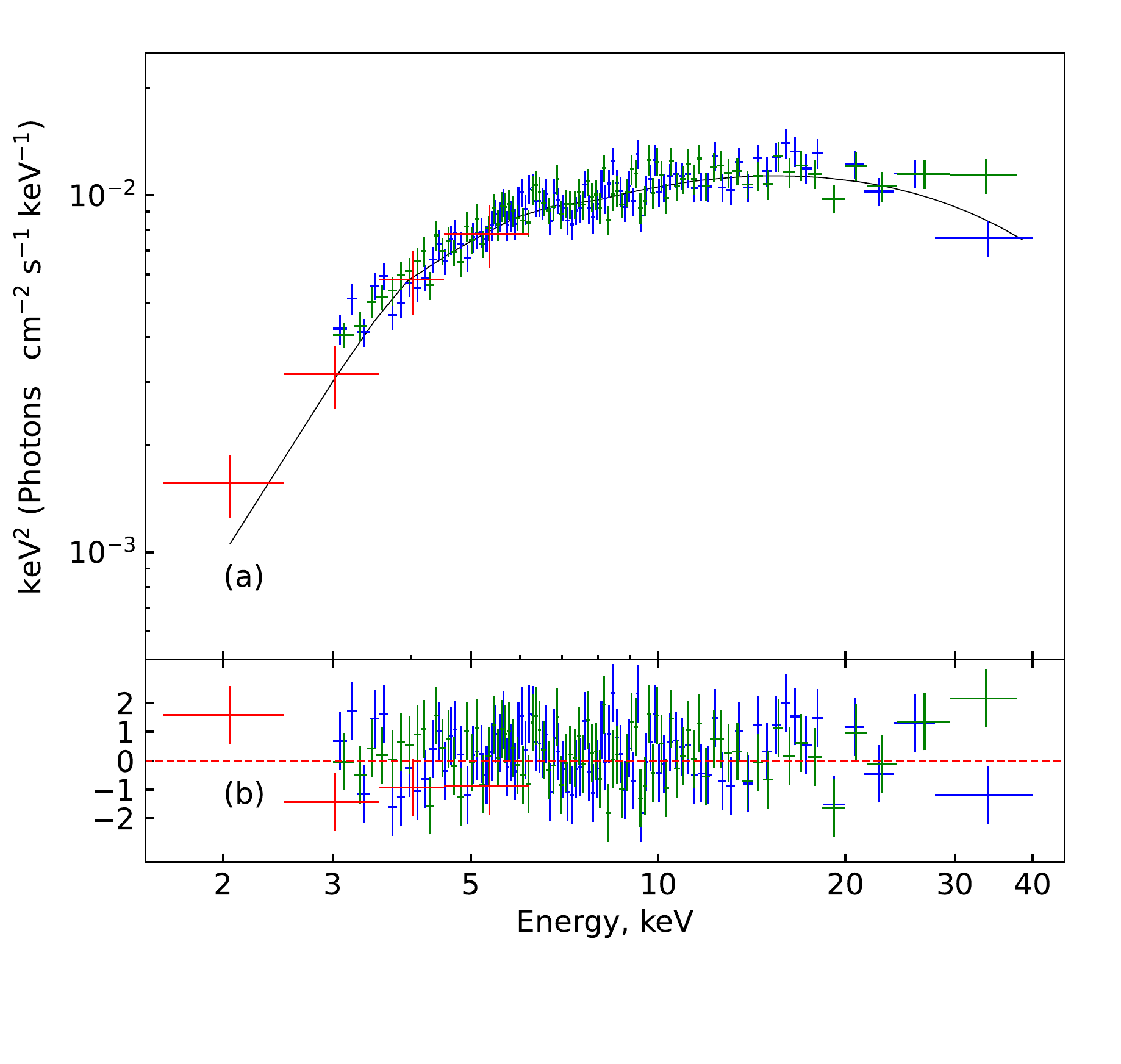}}
	\caption{(a) The energy spectrum of IGR~J21343+4738 from the NuSTAR data in December 2020 (ObsID 90601339002) (the green and blue dots are for the FPMA and FPMB modules, respectively) and the Swift/XRT data (red dots). The solid line indicates the best-fit model for the spectrum. (b) The deviation of the observational data from the POWERLAW*HIGHCUT model.} 
	\label{fig:spec}
\end{figure*}

The average energy spectrum of IGR~J21343+4738 from the NuSTAR and Swift/XRT data is shown in Fig. \ref{fig:spec}~a.  To describe the continuum of the source, we applied several models that are used to describe the spectra of X-ray pulsars in the XSPEC~v12.12.1 package \citep{1999ascl.soft10005A}. 
To compare our results with the previous study by \cite{2014A&A...561A.137R}, we chose a continuum model in the form of a power law with a high-energy exponential cutoff {\sc (powerlaw*highcut)}, which also demonstrated the best quality of the fit. To take into account the interstellar absorption, we added the {\sc tbabs} component to the model. The obtained absorption $N_{\rm H} \simeq 2.97\times 10^{22}$ cm$^{-2}$  exceeds the Galactic absorption toward the source $\sim 3.24\times 10^{21}$ cm$^{-2}$ \citep{2016A&A...594A.116H}, suggesting the presence of additional absorption in the system itself. \cite{2014A&A...561A.137R} used a model with the addition of a component in the form of blackbody radiation.
However, since its temperature is 0.11 keV, there is no need to include this component in our model due to the insufficient statistics in the soft energy bands. 
To take into account the possible differences in the effective area calibrations for the NuSTAR (FPMA, FPMB) and Swift/XRT instruments and the incomplete simultaneity of the source’s observations by different observatories, we introduced normalization coefficients (see Table \ref{tbl_1}). 
The remaining model parameters for different instruments were fixed between themselves. As in our timing analysis, the average spectrum is statistically significant up to 40 keV and, therefore, the parameters given below were obtained in the energy range 3–40 keV.

The observed absorbed X-ray flux from the source IGR J21343+4738 in the energy range 3–40 keV was $F_{x} = 3.77_{-0.53}^{+0.09} \times 10^{-11}$ erg cm$^{-2}$ s$^{-1}$1, corresponding to a luminosity $L_{x} \simeq 3.3\times 10^{35}$ erg s$^{-1}$ for a distance of 8.5 kpc.

As has already been said, the {\sc tbabs*powerlaw*highecut} model, 
which showed a quality of the fit equal to 553.50 per 538 degrees of freedom, 
gave the best fit. The model parameters are presented in Table \ref{tbl_1}. 
To test the hypothesis about the presence of a cyclotron feature in the source’s spectrum, we applied the approach proposed by \cite{2005AstL...31..380T}. In the energy range 3–40 keV the gabs component from the XSPEC package, an absorption line with a Gaussian profile, was added to the fitting model. The cyclotron line energy was changed in the specified energy range with a 5 keV step, while its width was fixed at 1 keV for line energies below 10 keV and at 2 keV for line energies above 10 keV.
The statistics improvement significance was tested using the $\chi$-square test. Thus, in the above energy range we detected no cyclotron lines whose inclusion would improve the quality of the spectral fit by more than 2$\sigma$, with the upper limit on the line optical depth being $\tau=0.18$. 

This result may suggest that the cyclotron line energy lies outside the energy range being investigated (3–40 keV), since at higher energies the statistics is not enough for the detection of a cyclotron line in the source’s spectrum. Thus, based on the relation for the cyclotron line energy B~=~$\frac{E_{cycl} (1 + z)}{11.6}\times 10^{12}$ G \citep{2019A&A...622A..61S}, we can assume that the magnetic field on the neutron star surface near the poles must be either less than $2.5 \times 10^{11}$ G or greater than $3.4 \times 10^{12}$ G. 

\begin{table}[ht]
\caption{Parameters of the spectrum of  \igr\ for the Powerlaw*Highecut model}
\centering
\label{tbl_1}
\begin{tabular}{cccc}	\\
\hline 
Model parameters &  \\ [0.5ex] 
\hline
$N_{H}$, {$10^{22}$ cm$^{-2}$} &   ${2.92_{-1.03}^{+1.13}}$ \\

$\Gamma$ & ${1.34\pm{0.13}}$ \\
$E_{cutoff}$, keV  &  ${5.9\pm{0.8}}$ \\
$E_{Fold}$, keV  &  ${22.4_{-3.9}^{+6.3}}$ \\
$Flux$ (3 - 40 keV), $10^{-11}$ &$3.77_{-0.3}^{+0.04}$ \\
erg cm$^{-2}$ s$^{-1}$& & \\

$C_{NuSTAR, FPMB}$&$1.01\pm{0.02}$ \\
$C_{Swift, XRT}$&$0.86_{-0.14}^{+0.15}$ \\
\hline
$\chi^{2}$ (d.o.f.) & 553.50(538) \\
\hline
&
\end{tabular}
\end{table}

Taking into account the fact that for a number of X-ray pulsars there are no cyclotron features in the averaged spectrum of the source, but, at the same time, they are detected at individual neutron star rotation phases \citep{2019ApJ...883L..11M, 2021ApJ...915L..27M}, we also carried out phase-resolved spectroscopy.  To  search for possible features in the spectra, depending on the rotation phase, we divided the data into four phase intervals corresponding to the rising pulse with a small peak (0–0.25), the primary peak (0.25–0.5), the decaying pulse (0.5–0.75), and the minimum pulse (0.75–1.0) in the light curve. The spectra of the four phase intervals were fitted by the same {\sc tbabs*powerlaw*highecut}model as that used in analyzing the averaged spectrum. The derived model parameters of the phase-resolved spectra hardly change with phase and have values approximately equal to those of the averaged spectrum. There are no features associated with the possible presence of a cyclotron absorption line.

\subsection{Orbital Period of the Binary System}

\begin{figure*}[!t]
	\center{\includegraphics[scale=0.78]{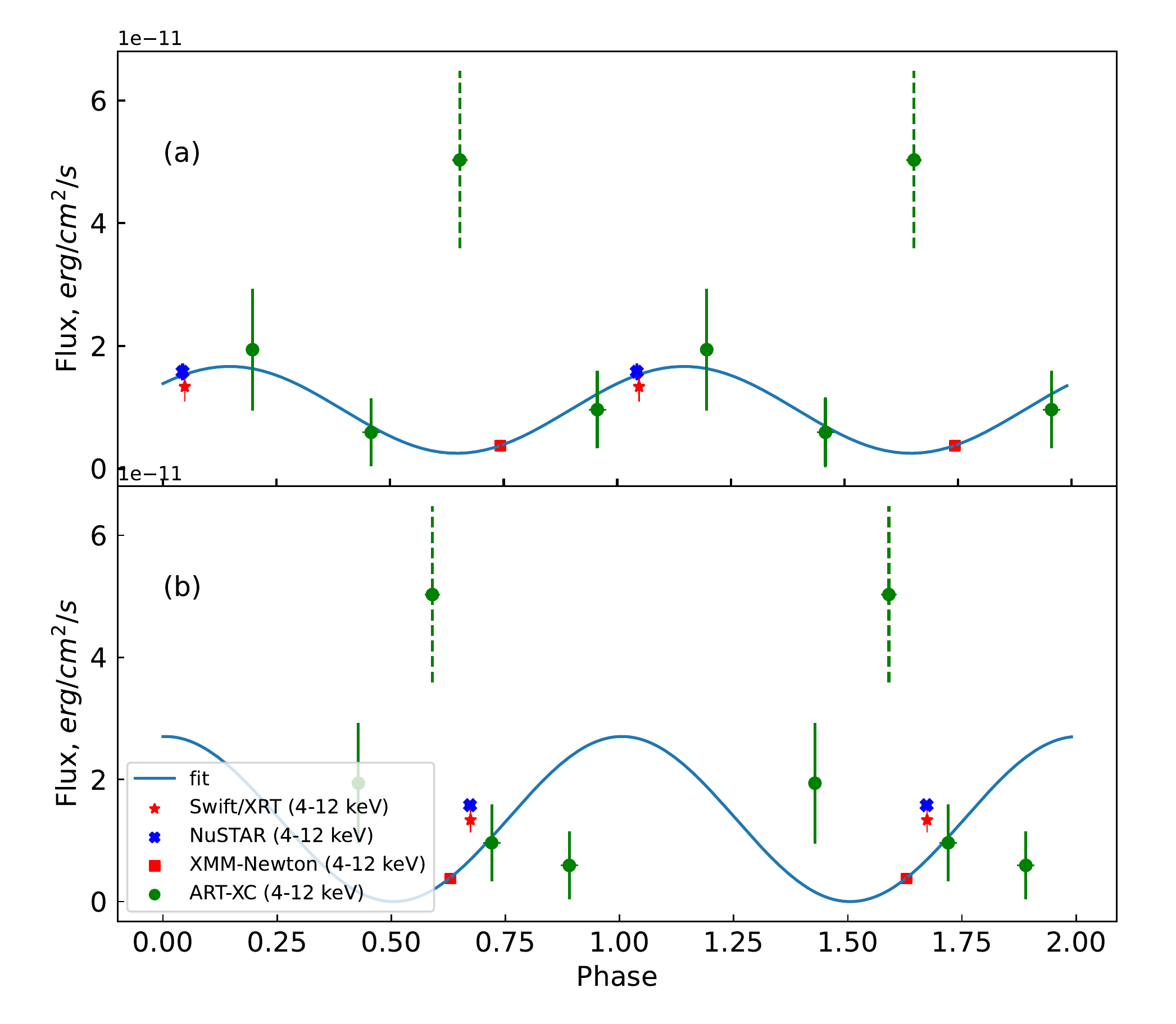}}
	\caption{The light curves of the X-ray pulsar IGR~J21343+4738 from the data of several telescopes folded with a period of 34.26 (a) and 160.8 (b) days. The fluxes from the source at phases 0.65 (a) and 0.59 (b) (marked by the dashed line) were obtained in December 2020 during a suspected outburst; therefore, this data point was excluded from the data series in our fitting.}
	\label{fig:orbit_lc}
\end{figure*}

\cite{2021mobs.confE..25N} determined two possible orbital periods of the neutron star around its companion, 34.26 and 160.8 days. To test these assumptions, we collected all of the available flux measurements for the source with the instruments onboard the SRG (ART-XC), Swift (XRT), NuSTAR, and XMM-Newton observatories and constructed long-term light curves in the 4–12 keV energy band. Then, they were folded with the above periods and fitted by the corresponding sine waves (Fig. \ref{fig:orbit_lc}). It can be seen from the figure that all of the flux measurements are acceptably described by the sine wave with the period of 34.26 days, except for the data point at phase 0.65. The corresponding measurement was carried out in December 2020 during the source’s suspected outburst and, therefore, this data point was excluded from the data series in our further fitting. The model describes the light curve with an acceptable statistic, $\chi^2/d.o.f.$~=~1.6/3, for the period of 34.26 days. The second model, with the period of 160.8 days, describes the data much more poorly, with a statistic $\chi^2/d.o.f.$~=~48.7/3. Thus, the available X-ray data are consistent with the assumption about the existence of an orbital period $\simeq34.3$ days in the system. However, long-term observations with X-ray instruments aimed at determining the changes in the source’s pulsation period associated with the orbital motion are needed for the final conclusions to be reached.  
  
\section{CONCLUSIONS}

In this paper we analyzed the observational data for the X-ray pulsar IGR~J21343+4738 obtained by the NuSTAR observatory and the Swift/XRT and SRG/ART-XC telescopes. We showed that the broadband spectrum of the source measured in December 2020 in a state with a luminosity $L_x \simeq 3.3 \times 10^{35}$ erg s$^{-1}$ could be best described by the continuum model in the form of a power law with a high-energy exponential cutoff {\sc powerlaw*highecut}. The component to take into account the interstellar absorption that exceeds the Galactic absorption toward the source, suggesting that part of the emission is absorbed inside the binary system, was included in the model. We also performed phase-resolved spectroscopy, which showed that the model parameters hardly change with phase.

Our timing analysis revealed X-ray pulsations with a period $\simeq322.71$ s. This value is greater than the one measured in 2013 based on XMM-Newton data by almost 2.5 s, suggesting a neutron star spin down with a mean rate  $\dot P/P \simeq 1\times10^{-3} $ yr$^{-1}$.  

The pulse profile is typical for persistent X-ray binaries with low fluxes \citep[see, e.g.,][]{1999MNRAS.306..100R}. In the energy range 3–40 keV it hardly changes with energy and agrees well with the previously measured one in softer channels. The pulsed fraction has the form of an increasing dependence on energy typical for X-ray pulsars.

No features that could be interpreted as cyclotron lines were detected in the spectrum of the source at energies 3–40 keV. Thus, we can assume that the magnetic field of the source is $B < 2.5\times 10^{11}$ G or $B > 3.4 \times 10^{12}$ G. 

The available measurements of the X-ray flux from the source in the 4–12 keV energy band are consistent with the assumption about the existence of an orbital period $\simeq34.3$ days in the system, but long-term observations are needed for the final conclusions to be reached. Assuming that the pulsar is close to the equilibrium state, and the accretion proceeds in the Bondi quasi-spherical subsonic accretion regime \citep{2012MNRAS.420..216S}, the magnetic field can be independently estimated by the orbital and rotational periods and the average accretion rate. Such an estimate gives the value $B \sim 2 \times 10^{11}$ G for the characteristic wind speed of massive systems $\sim$1 thousand km/s, which is consistent with the upper limit of magnetic field obtained from spectral analysis.

\section{ACKNOWLEDGMENTS}

We used the data obtained with NuSTAR, a Caltech project, funded by NASA and operated by NASA/JPL and the data provided by the UK Swift Science Data Centre (XRT data analysis). 
We also used the software provided by the High-Energy Astrophysics Science Archive Research Center (HEASARC). 
We used data from the Mikhail Pavlinsky ART-XC telescope onboard the SRG observatory.  The SRG observatory was built by Roskosmos in the interests of the Russian Academy of Sciences within the framework of the Russian Federal Space Program, with the participation of the Deutsches Zentrum f\"{u}r Luftund Raumfahrt (DLR).  The Mikhail Pavlinsky ART-XC telescope was built by the Space Research Institute of the Russian Academy of Sciences in collaboration with the Russian Federal Nuclear Center (Sarov, Russia). The Marshall Space Flight Center, NASA (USA), also participates in the project.  
The SRG spacecraft was designed, built, launched and is operated by the Lavochkin Association. The science data are downlinked via the Deep Space Network Antennae in Bear Lakes, Ussurijsk, Evpatoria, and Baykonur, funded by Roskosmos.  This work was supported by RSF grant no. 19-12-00423.

\bibliographystyle{mnras}
\bibliography{citations} 

\end{document}